\documentclass{sig-alternate}
\usepackage{mathptmx} 

\usepackage[sort,nocompress]{cite}
\usepackage{amsmath,amssymb,amsfonts}
\usepackage{algorithmic}
\usepackage{textcomp}
\usepackage{graphicx}
\usepackage[dvipsnames]{xcolor}
\usepackage{verbatim}
\usepackage{comment}
\usepackage{csquotes}
\usepackage{float}
\usepackage{multirow}
\usepackage{array}
\usepackage{makecell}
\usepackage{caption}
\usepackage{subcaption}
\usepackage{hhline}
\usepackage{pifont}
\usepackage{booktabs}
\usepackage{authblk}
\usepackage{fancyhdr}
\usepackage[normalem]{ulem}
\usepackage[hyphens]{url}
\usepackage[final]{microtype}
\usepackage[keeplastbox]{flushend}
\usepackage{listings}
\usepackage[bookmarks=true,breaklinks=true,colorlinks,citecolor=blue,linkcolor=blue,urlcolor=blue]{hyperref}

\lstdefinestyle{mystyle}{
    basicstyle=\ttfamily\footnotesize,
    breakatwhitespace=false,
    breaklines=true,
    captionpos=b,
    keepspaces=true,
    numbers=none,
    numbersep=5pt,
    showspaces=false,
    showstringspaces=false,
    showtabs=false,
    tabsize=2
}
\newcolumntype{x}[1]{>{\centering\arraybackslash\hspace{0pt}}p{#1}}
\newcolumntype{C}[1]{>{\centering\arraybackslash\hspace{0pt}}p{#1}}
\newcolumntype{R}[1]{>{\raggedleft\arraybackslash\hspace{0pt}}p{#1}}
\newcolumntype{L}[1]{>{\raggedright\arraybackslash\hspace{0pt}}p{#1}}
\newcommand{\ra}[1]{\renewcommand{\arraystretch}{#1}}

\newcommand{\tss}[1]{\textsuperscript{#1}}
\newcommand{\blackparrot}{BlackParrot}
\newcommand{\bedrock}{BedRock}
\newcommand{\bpbedrock}{BP-BedRock}

\title{The \blackparrot{} \bedrock{} Cache Coherence System}

\author[1]{Mark Wyse}
\author[1]{Daniel Petrisko}
\author[2]{Farzam Gilani}
\author[1]{Yuan-Mao Chueh}
\author[1]{Paul Gao}
\author[2]{Dai Cheol Jung}
\author[2]{Sripathi Muralitharan}
\author[2]{Shashank Vijaya Ranga}
\author[1,2]{Mark Oskin}
\author[1,2]{Michael Taylor}
\affil[1]{Paul G. Allen School of Computer Science \& Engineering}
\affil[2]{Department of Electrical \& Computer Engineering}
\affil[ ]{University of Washington}
\affil[ ]{Seattle, WA, USA}

\begin{document}
\maketitle

\begin{abstract}

This paper presents \bpbedrock{}, the open-source cache coherence protocol and system implemented within the \blackparrot{} 64-bit RISC-V multicore processor. \bpbedrock{} implements the \bedrock{} directory-based MOESIF cache coherence protocol and includes two different open-source coherence protocol engines, one FSM-based and the other microcode programmable. Both coherence engines support coherent uncacheable access to cacheable memory and L1-based atomic read-modify-write operations.

Fitted within the \blackparrot{} multicore, \bpbedrock{} has been silicon validated in a GlobalFoundries 12nm FinFET process and FPGA validated with both coherence engines in 8-core configurations, booting Linux and running off the shelf benchmarks. After describing \bpbedrock{} and the design of the two coherence engines, we study their performance by analyzing processing occupancy and running the Splash-3 benchmarks on the 8-core FPGA implementations. Careful design and coherence-specific ISA extensions enable the programmable controller to achieve performance within 1\% of the fixed-function FSM controller on average (2.3\% worst-case) as demonstrated on our FPGA test system. Analysis shows that the programmable coherence engine increases die area by only 4\% in an ASIC process and increases logic utilization by only 6.3\% on FPGA with one additional block RAM added per core.

\end{abstract}

\section{Introduction} \label{sec:intro}

System designers and computer architects are leveraging open-source hardware to create new processors and systems at an increasingly rapid pace. Within this movement, the RISC-V\cite{riscv-spec} ISA has been a key disruptive technology and has opened the door for computer designers, from individuals to corporations, to innovate and provide unique, bespoke systems.

As part of this movement, the open-source, Linux-capable, cache-coherent, 64-bit RISC-V \blackparrot{} \cite{petrisko2020} multicore processor has been developed. \blackparrot{} advanced the domain of processor design by adopting a software engineering approach for hardware with agile development, rapid design iteration, and continuous verification and testing to ensure a high-quality, high-performance design. The \blackparrot{} multicore processor has been silicon validated in a GlobalFoundaries 12-nm FinFET tapeout and FPGA validated on Xilinx Zynq and Xilinx Virtex Ultrascale+ HBM platforms \cite{xilinx-ultrascale}.

In this paper, we describe the implementation of \blackparrot{}'s coherence system, called \blackparrot{} \bedrock{} (\bpbedrock{}). We first provide a brief description of the \bedrock{} cache coherence protocol, a family of directory-based invalidate cache coherence protocols using the common MOESIF coherence states. \bedrock{} favors reducing protocol complexity rather than maximizing protocol concurrency and is well suited for small- (4-core) to medium-scale (32-core) multiprocessor designs.

We describe the implementation of \bpbedrock{}'s two directory-based coherence engines: (1) a directly implemented FSM, and (2) a microcode-programmable coherence engine. The microcode engine executes system firmware implementing one of the \bedrock{} coherence protocols. The protocol code can be extended with custom system- and application-specific functionality and allows experimentation with new features post-fabrication.

The two coherence engine designs of \bpbedrock{} are analyzed and evaluated to understand the performance and complexity tradeoffs involved in introducing programmability into the coherence engine. We present request processing occupancy for each coherence engine and discuss how to accelerate request processing with coherence-specific ISA extensions. We provide area overheads from an ASIC implementation of \bpbedrock{} and show the programmable coherence engine increases die area by only 4\% relative to a design utilizing the FSM-based engine. Analysis of an FPGA implementation shows the programmable engine increases logic LUT resource utilization by only 6.3\% alongside one additional block RAM used per programmable engine. We evaluate the system-level performance of both coherence engine designs using the Splash-3 benchmark suite running atop a Linux-based OS on the \bpbedrock{} FPGA implementations, and show that the programmable coherence engine has a performance overhead of only 1\% on average and 2.3\% for the worst-case benchmark. These results encourage the exploration of programmability within the cache coherence engine and unique system features programmability can unlock.

The rest of this paper is organized as follows. Section \ref{sec:bedrock} briefly presents the \bedrock{} cache coherence protocol. Section \ref{sec:directory} describes the implementation \bpbedrock{}'s coherence directory storage. Sections \ref{sec:fsm_cce} and \ref{sec:ucode_cce} describe the two cache coherence engine (CCE) implementations available in \bpbedrock{}. Section \ref{sec:evaluation} presents the evaluation of the \bpbedrock{} system. Section \ref{sec:related} describes prior research related to \bpbedrock{}, and Section \ref{sec:conclusion} concludes.

\begingroup
\setlength{\tabcolsep}{3pt} 
\begin{table*}[t]\centering
\begin{tabular}{@{}L{0.05\linewidth}L{0.06\linewidth}L{0.06\linewidth}ccccccccC{0.1\linewidth}@{}}\toprule
& \multicolumn{2}{c}{Cache Action} & \phantom{a} & \multicolumn{8}{c}{Coherence Message}\\
\cmidrule{2-3} \cmidrule{5-12}
State & Load & Store && Inv & DATA & STW & WB & TR & ST-WB & ST-TR & ST-TR-WB\\
\midrule
I & ReqRd & ReqWr && & CohAck/X & & & & & &\\
S & Hit & ReqWr && InvAck/I & & CohAck/M & & & & &\\
E & Hit & Hit/M && & & & NullWB/E & & NullWB/X & DATA/X & DATA, NullWB/X\\
M & Hit & Hit && & & & DirtyWB/M & & DirtyWB/X & DATA/X & DATA, DirtyWB/X\\
O & Hit & ReqWr && & & CohAck/M & DirtyWB/O & DATA/O & DirtyWB/X & DATA/X &\\
F & Hit & ReqWr && & & CohAck/M & & DATA/F & & DATA/X &\\
\bottomrule
\end{tabular}
\caption{\bedrock{} Cache Controller Protocol Table - MOESIF. X state indicates a valid state provided by coherence directory. DATA messages are sent to another controller over the \bedrock{} Fill network.}
\label{table:bedrock-protocol-lce-moesif}
\end{table*}
\endgroup

\begin{table*}[t]\centering
\begin{tabular}{@{}L{0.05\linewidth}L{0.13\linewidth}L{0.13\linewidth}L{0.14\linewidth}L{0.14\linewidth}L{0.14\linewidth}L{0.12\linewidth}@{}}\toprule
\multirow{2}{*}{\parbox{0.05\linewidth}{Dir State}} & \multicolumn{5}{c}{Coherence Request} & \\
\cmidrule{2-6}
& ReqRd & ReqRd-NE & ReqWr from I & ReqWr from S & ReqWr from O/F & Replacement\\
\midrule
I & DATA to Req/E & DATA to Req/S & DATA to Req/M &&&\\
S & DATA to Req/S & DATA to Req/S & Inv all S, DATA to Req/M & Inv other S, STW\tss{M} to Req/M &&\\
E & ST\tss{F}-TR\tss{S}-WB to Owner/F & ST\tss{F}-TR\tss{S}-WB to Owner/F & ST\tss{I}-TR\tss{M} to Owner/M & && ST\tss{I}-WB to Req/I\\
M & ST\tss{O}-TR\tss{S} to Owner/O & ST\tss{O}-TR\tss{S} to Owner/O & ST\tss{I}-TR\tss{M} to Owner/M &&& ST\tss{I}-WB to Req/I\\
O & TR\tss{S} to Owner/O & TR\tss{S} to Owner/O & Inv all S, ST\tss{I}-TR\tss{M} to Owner/M & Inv other S and Owner, STW\tss{M} to Req/M & Inv all S, STW\tss{M} to Req/M & ST\tss{I}-WB to Req/I\\
F & TR\tss{S} to Owner/F & TR\tss{S} to Owner/F & Inv all S, ST\tss{I}-TR\tss{M} to Owner/M & Inv other S and Owner, STW\tss{M} to Req/M & Inv all S, STW\tss{M} to Req/M &\\
\bottomrule
\end{tabular}
\caption{\bedrock{} Coherence Directory Protocol Table - MOESIF. Superscript states are states attached to command messages.}
\label{table:bedrock-protocol-cce-moesif}
\end{table*}

\section{\bedrock{}} \label{sec:bedrock}

The \bedrock{} Cache Coherence Protocol \cite{bedrock} defines a family of directory-based invalidate cache coherence protocols using the common MOESIF cache coherence states and the coherence system components required to implement the protocol. This section provides a brief overview of the \bedrock{} protocol, which the \bpbedrock{} system described in this paper implements. We refer readers to the protocol specification \cite{bedrock} for additional details on the design decisions of the protocol itself.

A \bedrock{} system comprises one or more cache controllers (Local Cache Engines), one or more coherence directories (Cache Coherence Engines), and the coherence message networks. Each cache controller manages a single cache participating in the coherence protocol. The coherence directory is a standalone, inclusive, duplicate tag directory and acts as the point of serialization for all coherence transactions. Coherence is enforced using the \emph{Single-Writer, Multiple-Reader (SWMR) Invariant} and \emph{Data-Value Invariant} \cite{mcm-primer-v2}. \bedrock{} works with ordered or unordered networks and assumes the network implementation provides error free message delivery.

\bedrock{} differs from a canonical directory protocol in a few subtle ways. First, the coherence directory has complete control over all changes to cache block coherence states, including invalidation and eviction of blocks from the cache controllers. The sole exception is a cache may silently upgrade a block from the Exclusive (E) state to the Modified (M) state on a write operation. Second, \bedrock{} utilizes four unidirectional coherence networks, ordered in priority from highest to lowest: Response, Fill, Command, and Request. The Request and Response networks carry messages from cache controller to coherence directory, the Command network carries messages from directory to cache, and the Fill network carries cache to cache data transfers. Third, the cache controllers never hold a block in a transient state; all transient state is hidden from the controllers by the coherence directory's processing flow. Lastly, the protocol design favors reducing complexity over maximizing concurrency.

\subsection{Request Processing}

The canonical processing flow for a \bedrock{} request is divided into two stages. First, the directory is read and processed to determine if a cache block eviction (replacement) or invalidations are required. Then, the coherence engine determines the source of the cache block and initiates a cache to cache transfer, performs a memory read, or responds to the requesting cache with read/write permissions for the block if it already has a valid copy. The transaction completes when the cache controller receives the block and responds to the directory with a coherence acknowledgment. Advanced directory implementations may be able to overlap some of these actions or perform speculative memory accesses to reduce request processing latency.

\subsection{Protocol Tables}

Tables \ref{table:bedrock-protocol-lce-moesif} and \ref{table:bedrock-protocol-cce-moesif} present the complete tabular specification \cite{sorin2002} of the \bedrock{} MOESIF coherence protocol for the cache controller and coherence directory, respectively. Each table uses an "Action/State" notation to describe the behavior of the controllers. Given the current coherence state for a cache block (row) and an event (column), an entry in the table describes the action taken by the controller in response to the event and the next coherence state of the block at the controller. Blank entries indicate impossible state and event pairs for the controller.

\subsubsection{Cache Controller Protocol Table}

Table \ref{table:bedrock-protocol-lce-moesif} describes how the cache controller responds to cache actions (load or store) and coherence commands. Cache actions may either hit in the cache or trigger a new coherence request. Coherence commands are directives issued by the directory to modify the state of a block and send a message in response on the Response or Fill network. An X state indicates a valid coherence state provided by the directory in the command message that is not known \textit{a priori} by the cache controller.

\subsubsection{Coherence Directory Protocol Table}

Table \ref{table:bedrock-protocol-cce-moesif} describes how the coherence directory processes coherence requests. The directory may also evict cache blocks from a controller to make room for a newly requested block (Replacement). Each "Action/State" entry describes the command messages sent by the directory to complete the request and the next state of the block at the directory. Some commands generate a response to the directory, and all transactions are finalized when the cache controller sends a coherence acknowledgment to the directory. The coherence state superscripts attached to some messages provide an associated coherence state for the message. For example, an ST\tss{I}-TR\tss{M} message directs a cache controller to set the state of the target block to Invalid (I) and transfer that block in the Modified (M) state to another controller.

\subsection{Ordering Transactions - Way Groups} \label{sec:bedrock_way_groups}

\begin{figure}[t]
	\centering
	\includegraphics[width=\linewidth]{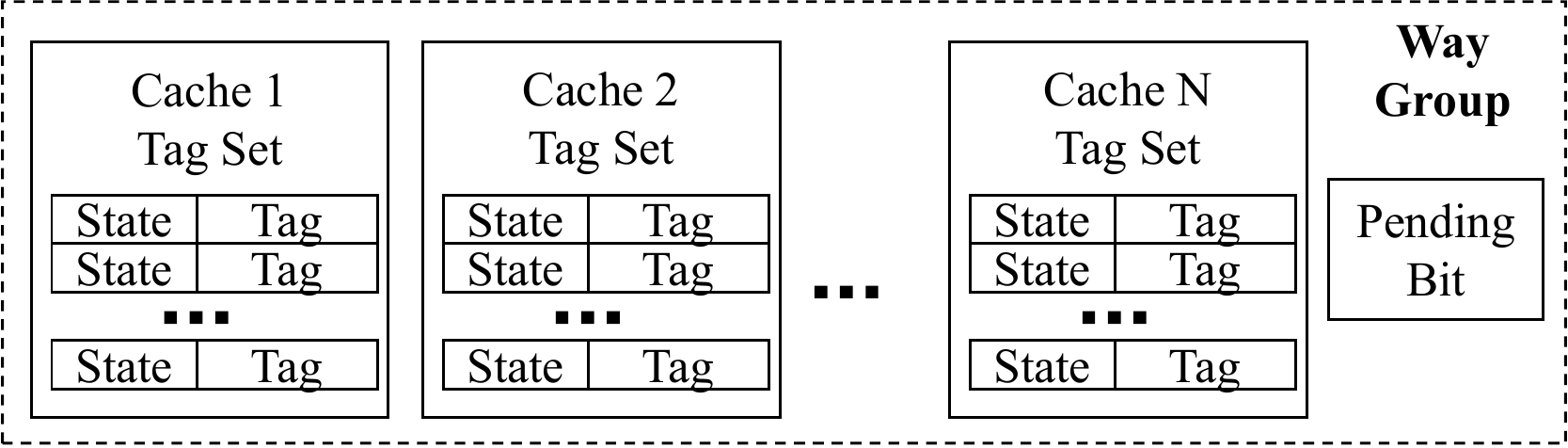}
	\caption{Way Group Organization}
	\label{fig:bedrock-way-group}
\end{figure}

The \bedrock{} protocol allows concurrent coherence transactions to independent cache blocks, but only one transaction at a time per group of related blocks. Two cache blocks are related if they belong to the same \emph{Way Group}. A way group, shown in Figure \ref{fig:bedrock-way-group}, is the collection of cache blocks that map to the same cache set in \emph{any} cache. A \emph{Tag Set} is simply the pairs of coherence state and address tag for all ways within a single set of a single cache. The \emph{Pending Bit} is set when a new transaction targeting a block in the way group starts and cleared when the active transaction completes. Transactions to way groups with a set pending bit must stall until the previous transaction completes. Requests targeting the same way group may be concurrently issued by many cache controllers and are serialized by the directory.

Every way group in the system is managed exclusively by a single coherence directory. In a system where all caches participating in coherence have the same organization with S sets, there are S way groups (one per cache set). A system with varied cache organizations has a number of way groups equal to the minimum number of cache sets across all caches.

Way groups guarantee that a coherence request will only cause changes to cache blocks within the target way group. Consequently, two transactions to two different way groups are guaranteed to be independent and can be processed concurrently. The use of way groups sacrifices protocol concurrency, but also eliminates transient states at the cache controllers that are typically required to handle that concurrency, thus simplifying the protocol for the cache controllers.\footnote{Future work will investigate efficient mechanisms to support concurrency within way groups.}

\section{\bpbedrock{} Directory} \label{sec:directory}

\begin{figure}[t]
	\centering
	\includegraphics[width=\linewidth]{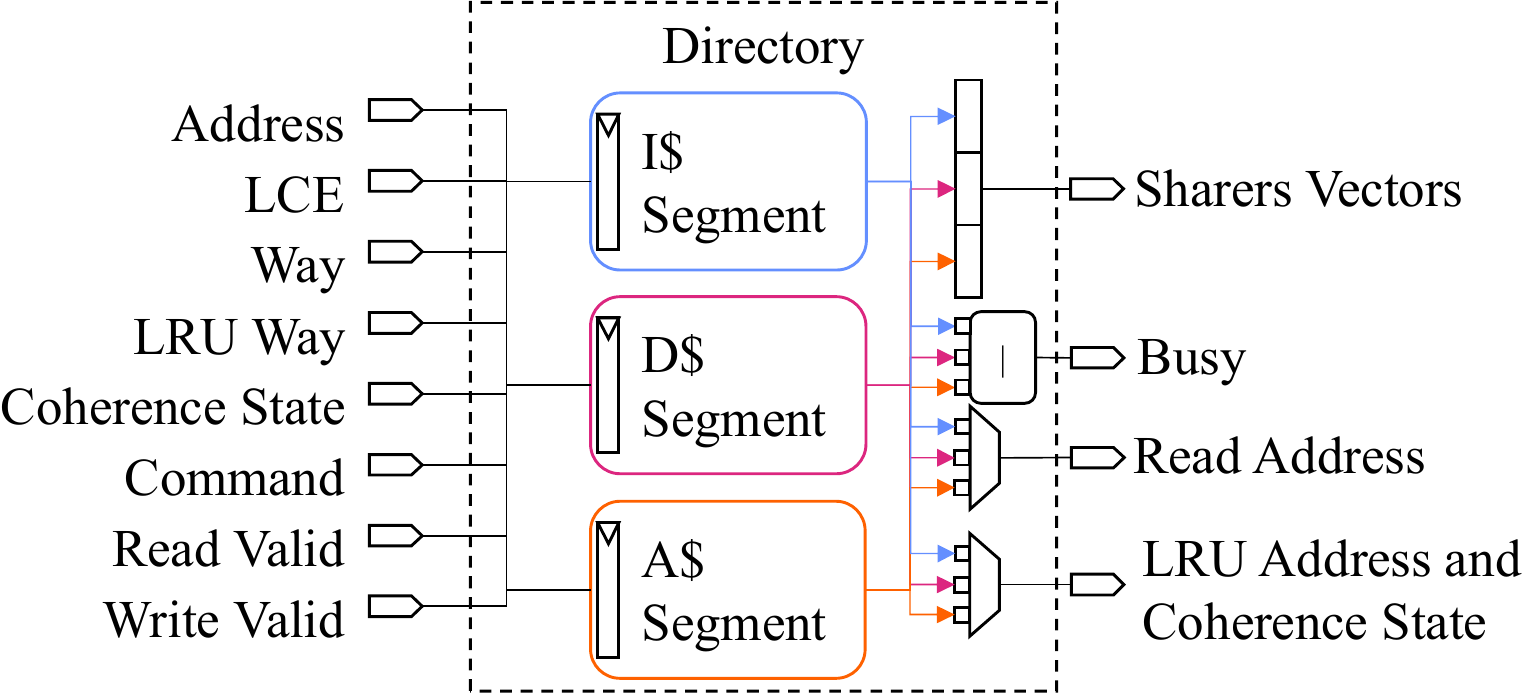}
	\caption{\bpbedrock{} Coherence Directory Architecture}
	\label{fig:cce-dir}
\end{figure}

The \bpbedrock{} cache coherence engines (CCE) implement complete standalone duplicate tag directories. The organization of the coherence directory is identical in both the hardware-based (FSM) and microcode programmable (ucode) coherence engines. The directory is constructed from multiple directory segments with one segment per cache type in the system. \bpbedrock{}'s cache types are instruction, data, and optional coherent accelerator caches. Figure \ref{fig:cce-dir} shows a block diagram of the coherence directory. Systems without coherent accelerators do not instantiate the accelerator cache segment.

\subsection{Coherence Directory Segment} \label{sec:directory_segment}

\begin{figure}[t]
	\centering
	\includegraphics[width=0.85\linewidth]{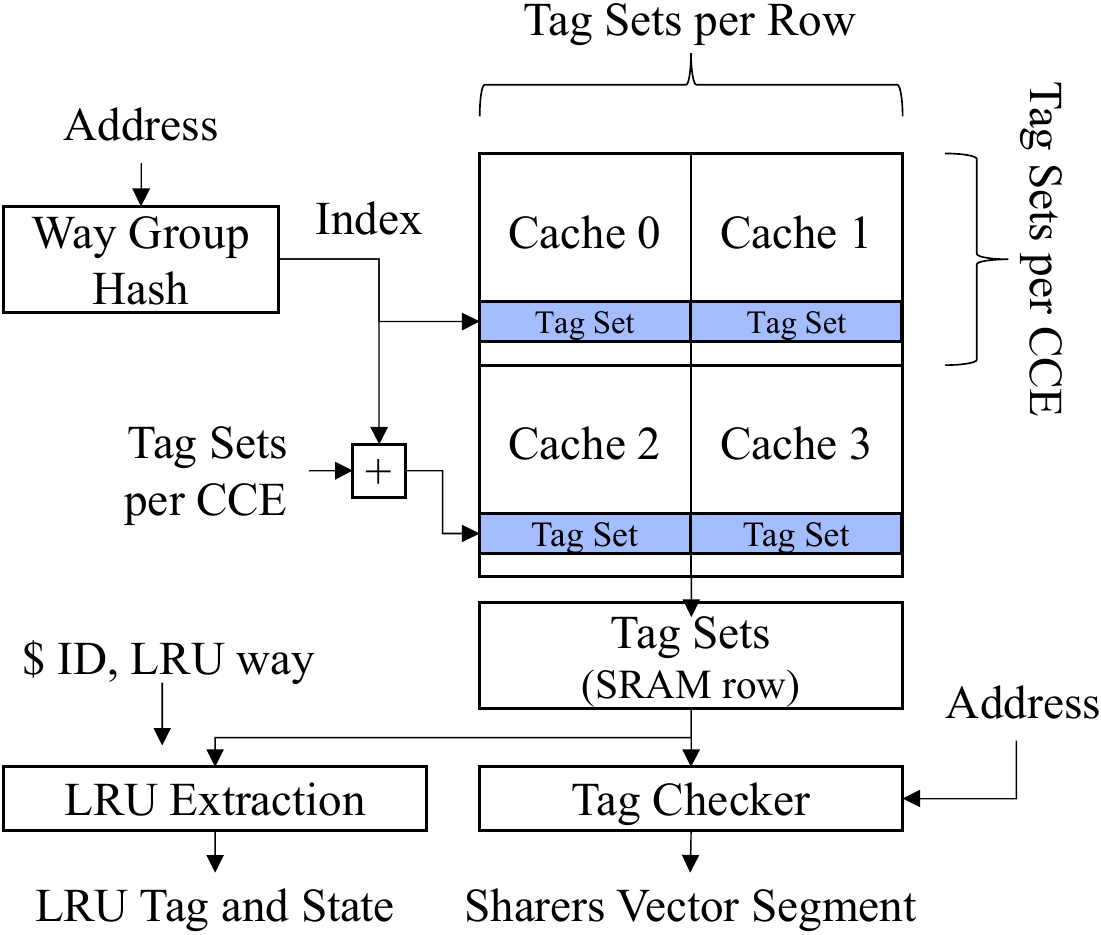}
	\caption{\bpbedrock{} Coherence Directory Segment}
	\label{fig:cce-dir-segment}
\end{figure}

Figure \ref{fig:cce-dir-segment} shows the organization of a \bpbedrock{} directory segment. Each segment stores a subset of the tag sets from all caches of a single type. The tag sets of all caches are spread evenly across the CCEs, with all tags sets in a single way group mapped to a single CCE. The tag sets of a single cache are stored in sequential rows in the directory SRAMs, and each row stores related tag sets from one or more caches. If the number of \emph{Tag Sets per Row} is less than the number of caches being tracked by the directory segment, additional blocks of \emph{Tag Sets per CCE} rows are added to track all caches.

The directory segment SRAM is a single-ported synchronous read-write memory. The width of the SRAMs are sized based on prior physical design work showing that two tag sets per row is PPA-efficient, however any power-of-two value is supported. The location of a cache block is easily computed using the cache's ID and the block address, with the cache ID providing the first row of the cache's tag set rows and the address providing a row offset within the block. The least significant bits of the cache's ID also provides a horizontal offset into the target row.

\subsection{Directory Operations and Access}

A small FSM controls each directory segment and supports tag set entry reads, way group reads of all tag sets in a way group, tag set entry writes, and physical row clears. The directory segment organization provides single cycle writes and multiple cycle reads. Write operations have a latency of one cycle, and a new write may be issued every cycle. Tag set entry reads require two cycles. Way group reads require two or more cycles, depending on the total number of caches being tracked by the segment. In a \bpbedrock{} multicore, a way group read requires $1 + (Cores/2)$ cycles for the default directory organization.

\subsection{Tag Checker and LRU Extraction}

\bpbedrock{} includes two modules that process the directory way group reads. The Tag Checker examines each directory row as it leaves the SRAM and produces the three Sharers Vectors. Each vector has one entry per cache, and the three vectors provide a cache hit bit, coherence state, and the cache way of the current LCE request's cache block. The LRU extraction module processes each directory row and extracts the tag set entry at the LRU way for the specified LCE. If the directory segment requires multiple rows to store all cache's tag sets for a single cache set, the LRU Extraction module outputs the LRU information only for the row containing the requesting LCE's tag set. The LRU way input is provided in the coherence request from the cache controller.

\section{\bpbedrock{} FSM CCE} \label{sec:fsm_cce}

\begin{figure}[t]
	\centering
	\includegraphics[width=.9\linewidth]{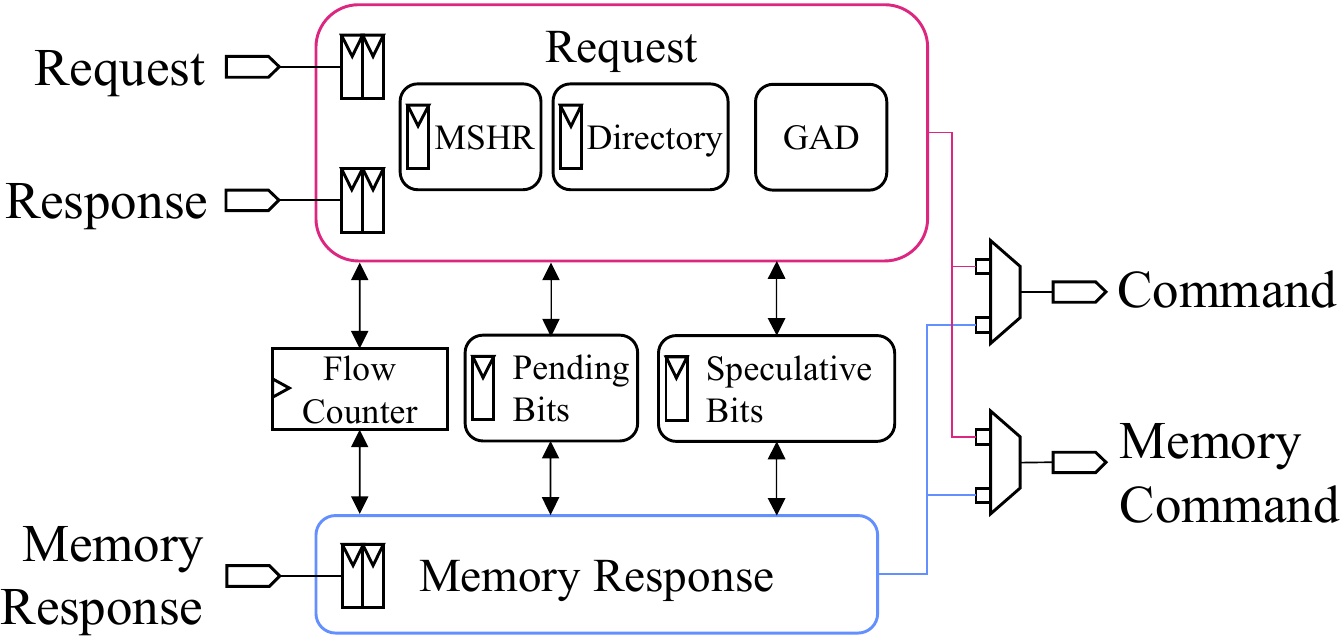}
	\caption{\bpbedrock{} FSM CCE Block Diagram}
	\label{fig:fsm-cce-bd}
\end{figure}

\bpbedrock{}'s FSM-based cache coherence engine (FSM CCE) is a direct hardware implementation of an FSM implementing the \bedrock{} MOESIF coherence protocol of Table \ref{table:bedrock-protocol-cce-moesif}. Figure \ref{fig:fsm-cce-bd} shows the FSM CCE's block diagram comprising Request and Memory Response state machines, Speculative Bits to track speculative memory reads, Pending Bits to enforce coherence transaction ordering for each way group, and a Flow Counter to provide memory network flow control.

\subsection{Request FSM}

The Request state machine processes coherence requests from the cache controllers. It instantiates a Miss Status Handling Register (MSHR) to hold processing state, the coherence directory, and a GAD module that processes the coherence directory output. The coherence directory implementation is described in Section \ref{sec:directory}. The MSHR and GAD modules are explained below, followed by a description of the state machine.

\subsubsection{MSHR} \label{sec:fsm_cce_mshr}

\begin{table}[t]\centering
	\ra{1.3}
	\begin{tabular}{@{}L{0.4\linewidth}L{0.55\linewidth}@{}}\toprule
		Flag & Description \\
		\midrule
		Write Not Read & Write request\\
		Uncached & Uncached request\\
		Non Exclusive & Non-Exclusive request\\
		Atomic & Atomic operation request\\
		Atomic No Return & Atomic no return request\\
		Cacheable Address & Request to cacheable memory\\
		Pending & Pending bit set on last read\\
		Cached Shared & Block cached in S at other LCE\\
		Cached Exclusive & Block cached in E at other LCE\\
		Cached Modified & Block cached in M at other LCE\\
		Cached Owned & Block cached in O at other LCE\\
		Cached Forward & Block cached in F at other LCE\\
		Replacement & Replacement required\\
		Upgrade & Request can be resolved with a permission upgrade\\
		\bottomrule
	\end{tabular}
	\caption{\bpbedrock{} MSHR Flags}
	\label{table:cce-fsm-mshr-flags}
\end{table}

The MSHR accumulates information related to the current LCE request during request processing. This information includes the request address, type, size, and requesting controller's ID; information about any required cache block replacement; the current state, owner, and cache way of the requested block, and a set of control flag bits. The MSHR information is filled primarily from the request message begin processed and the required coherence directory read.

Table \ref{table:cce-fsm-mshr-flags} lists the control flow flags stored in the MSHR. These flags come mostly from the current LCE request and the GAD module after a directory read, and are used to make efficient control flow decisions. The top six flags record properties of the current request. The Pending flag is set by the result of a pending bits read operation. The remaining flags are set after reading and processing the directory. The Replacement flag is set if a cache block replacement is required, and the Upgrade flag is set if the request can be resolved with a read/write permission upgrade to the requesting cache. The Cached \emph{State} flags are set if the requested cache block exists in the corresponding state in any cache other than the requester.

\subsubsection{GAD}

During request processing, the FSM CCE reads the coherence directory to extract the current state of the target block across all caches in the system. The raw directory data is processed by the tag checker and output as the sharers vectors. The GAD, or Generate Auxiliary Directory Information, module consumes the sharers vectors and LRU information and computes a subset of the MSHR control flags; the owner, location, and coherence state of the target block, if an owner exists; and the cache way of the block within the requesting cache, if present. The GAD module takes a single cycle to execute and is invoked in the cycle following the coherence directory read.

\subsubsection{Request FSM}

\begin{figure}[t]
	\centering
	\includegraphics[width=\linewidth]{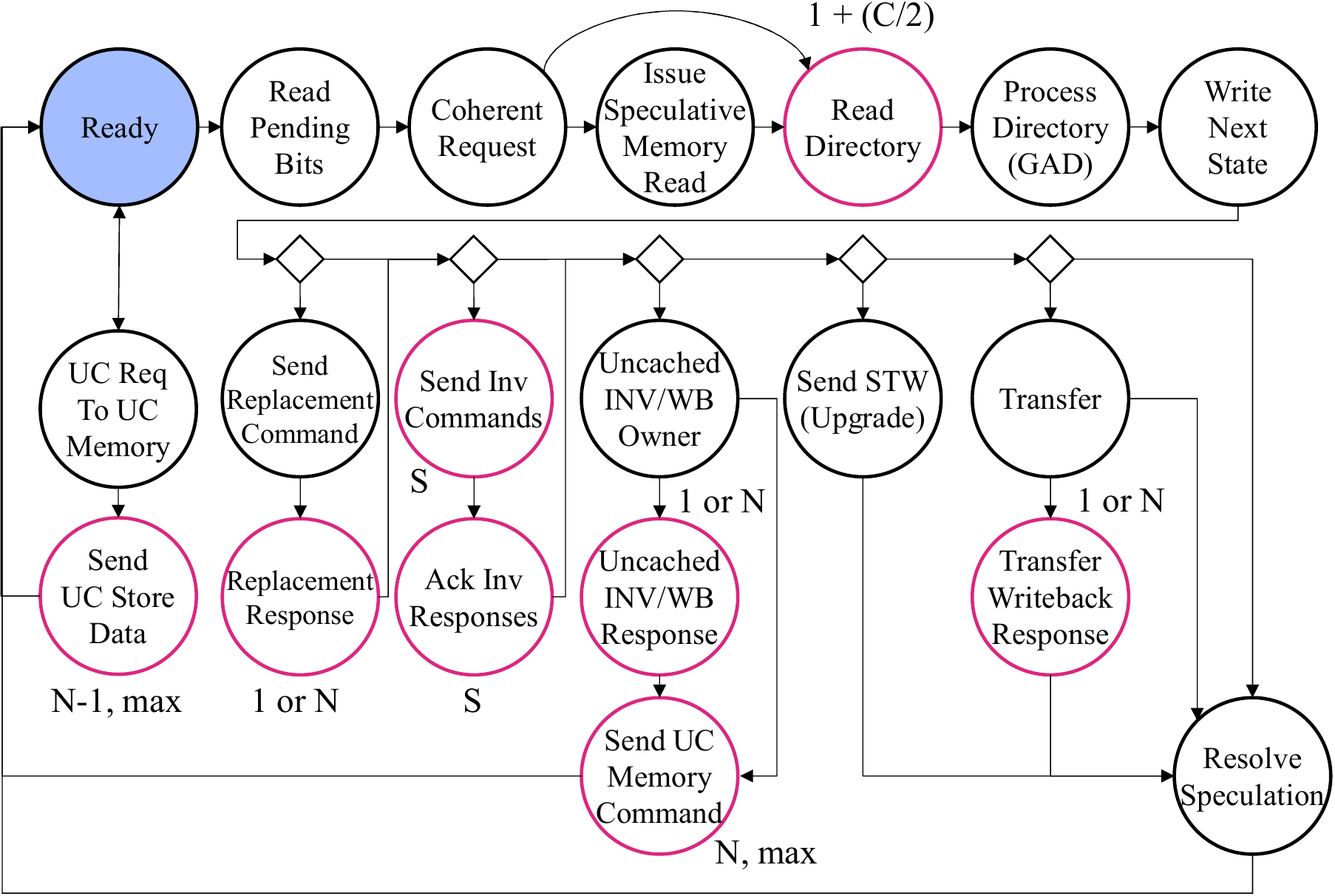}
	\caption{\bpbedrock{} FSM CCE Request State Machine}
	\label{fig:fsm-cce-req-fsm}
\end{figure}

Figure \ref{fig:fsm-cce-req-fsm} shows the FSM CCE's request processing state machine. The FSM processes one request at a time and runs without interruption. States requiring multiple cycles to execute (outlined in magenta) are labelled with an execution latency. All other states execute in a single cycle. Execution begins in the Ready state, where requests are classified by request address as either coherent if they target cacheable memory or uncacheable if targeting uncacheable memory.

All requests targeting cacheable memory participate in the coherence protocol and require a coherence directory read. These requests are processed through an initial set of states that check the pending bits for an open transaction to the same way group, issue a speculative memory read for cacheable requests, read and process the directory, and then write the next state for the block into the directory. After initial processing, the FSM executes only those steps required by the request. Diamonds in the state machine diagram represent control flow decisions and have a cost of zero cycles. A key benefit of the FSM-based CCE implementation is that it can concurrently perform protocol processing and make control flow decisions, thereby incurring no overhead to transition between protocol processing steps.

Cacheable requests may require a cache block replacement to evict a block from the requesting cache or cache block invalidations to caches that contain valid copies of the requested block. After replacement and invalidations are complete, the request is resolved by either sending an upgrade command when the requester needs read/write permissions and has a valid read-only copy of the block, initiating a cache to cache transfer if the block is owned by another cache, or confirming that the block will be sourced from memory.

Uncacheable requests to cacheable memory invalidate and writeback the target block from all caches that contain a valid copy, before issuing the uncached operation to memory.

Uncacheable requests to uncacheable memory require minimal processing and result in either an uncached load or store being issued to memory. Uncacheable requests require one cycle to send the message header and first data beat plus one cycle per additional data beat. These requests do not participate in the coherence protocol.

\subsubsection{Atomics} \label{sec:fsm_cce_atomics}
\bpbedrock{} supports atomics at the L1 and L2 caches, with L1 atomics enabled by default. L1 atomics result in the L1 data cache requesting write permission for the target block before the cache executes the operation after the block fill completes. RISC-V LR/SC sequences are handled similarly with the L1 data cache requesting write permissions at the LR operation, and are guaranteed to eventually succeed per the RISC-V specification. L2 atomic requests are executed at the L2 cache, with the coherence directory invalidating the target block from all L1 caches before the operation is sent to the L2 cache. All atomic operations are ordered by the coherence protocol (requesting write permissions for L1 atomic or LR/SC) or the coherence directory (serialization of requests for L2 atomic).

\subsection{Pending Bits}

The FSM CCE implements the pending bits attached to each way group as a collection of small counters external to the coherence directory. A way group's pending bit is considered set if the counter is non-zero and unset if the counter is zero. Pending bit writes are synchronous, while reads are asynchronous and support write to read forwarding. The CCE increments a way group's pending bit when beginning request processing or issuing a memory command. Pending bits are decremented by consuming memory responses or coherence acknowledgment responses, and when the CCE finishes processing uncached requests to coherent memory.

\subsection{Speculative Bits}

The Speculative Bits record information about speculative memory reads issued during request processing. There is one Speculative Bits entry per way group. Each entry includes a coherence state, a speculative bit, a squash bit, and a forward-modified bit. The CCE sets a way group's speculative bit when issuing a speculative memory request and clears the bit once the source of the cache block is determined after reading the directory. A speculative memory request will be squashed if the squash bit is set, forwarded with the stored coherence state when the forward-modified bit is set, or forwarded without modification if all bits have been cleared. Memory requests are squashed when the block can be supplied by a cache to cache transfer.

\subsection{Memory Response FSM}

The Memory Response FSM is a three-state multi-cycle state machine that processes \bpbedrock{} Memory Response messages returning from the L2 cache / memory or I/O devices. The state machine examines each memory response as it arrives and either forwards it to the appropriate LCE or sinks the response from the memory network. All memory responses include a speculative bit that is set if the response was generated by a speculative memory command. Speculative responses are either squashed or forwarded to a cache controller, as explained above. Non-speculative responses either carry data from a cached or uncached read, which are forwarded to the requesting controller, or are header-only responses to memory write commands, which are sunk by the FSM.

\section{\bpbedrock{} ucode CCE} \label{sec:ucode_cce}

\begin{figure}[t]
	\centering
	\includegraphics[width=0.98\linewidth]{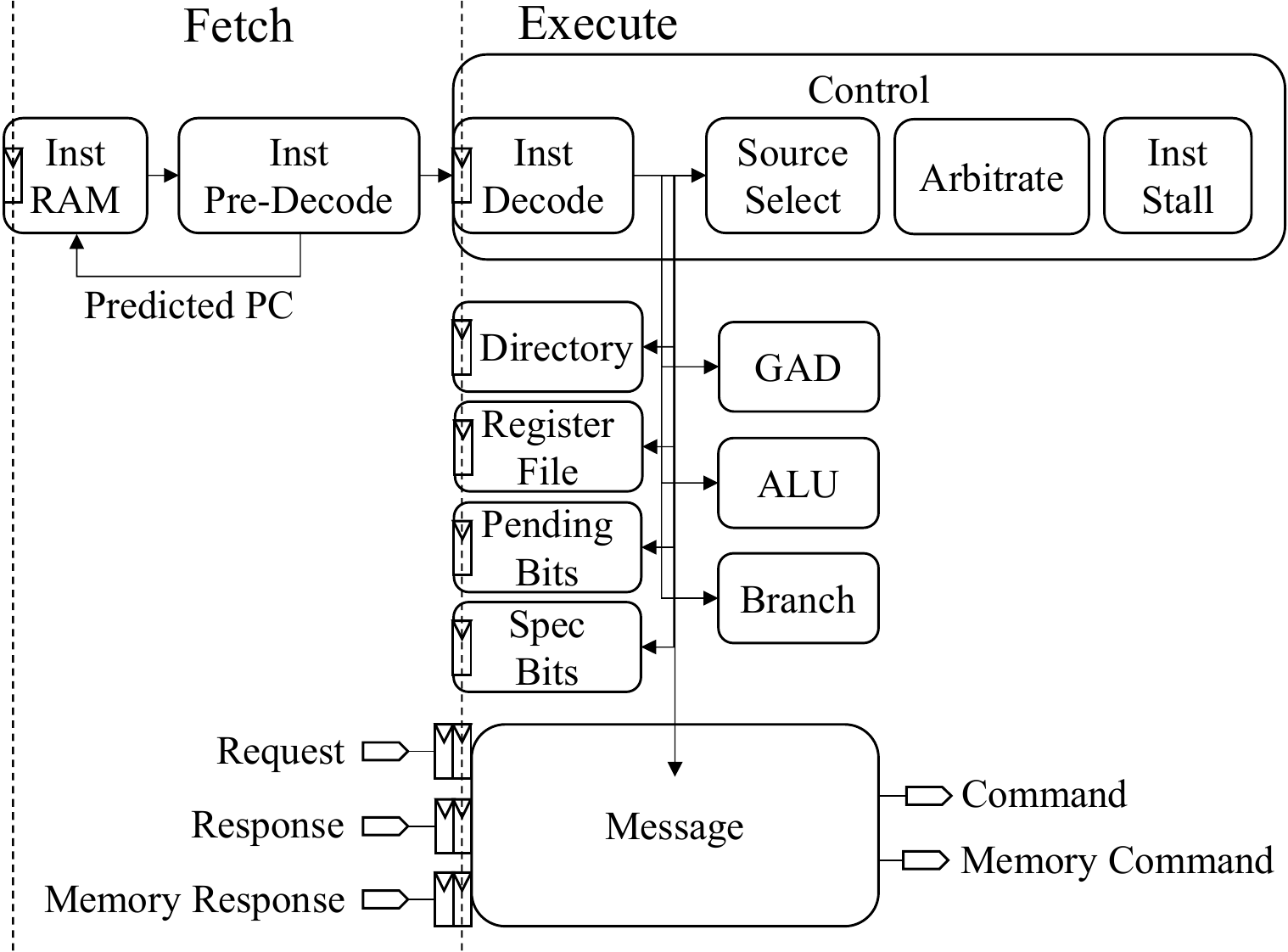}
	\caption{\bpbedrock{} Microcode-Programmable CCE Block Diagram}
	\label{fig:ucode-cce-bd}
\end{figure}

The \bpbedrock{} microcode-programmable CCE (ucode CCE) is a user-programmable coherence engine with a two-stage fetch-execute pipeline, 64-bit general purpose registers and datapath, specialized coherence protocol processing logic, and a custom RISC ISA specialized for coherence protocol processing. The ucode CCE is specialized for the \bedrock{} coherence protocol and provides programmability to enable system- and application-specific functionality on top of the coherence protocol. Figure \ref{fig:ucode-cce-bd} shows a block diagram of the \bpbedrock{} ucode CCE. The programmable CCE reuses many modules from the FSM CCE, including the Coherence Directory, GAD, Pending Bits, and Speculative Bits. \bpbedrock{} provides microcode implementations of the \bedrock{} MOESIF, MESI, MSI, and EI protocols. The baseline MOESIF protocol with speculative memory access closely matches the processing flow of the FSM CCE implementation. The remainder of this section describes the microcode ISA and modules unique to the ucode CCE.

\subsection{Microcode ISA} \label{sec:ucode_cce_isa}

\begin{table*}[t]\centering
	\begin{tabular}{@{}L{0.09\linewidth}L{0.54\linewidth}L{0.32\linewidth}@{}}\toprule
		Op & Format & Function\\
		\midrule
		sf, sfz & op flag & flag = 1, flag = 0\\
		andf, orf & op flag flag gpr & rd = flag op flag\\
		nandf, notf & op flag flag gpr & rd = flag op flag\\
		notf & notf flag gpr & rd = ~flag\\
		bf & bf tgt flag [flag...] [pt] & pc = tgt if all flags 1\\
		bfnot & bfnot tgt flag [flag...] [pt] & pc = tgt if all flags 0\\
		bfz & bfz tgt flag [flag...] [pt] & pc = tgt if any flag 1\\
		bfnz & bfnz tgt flag [flag...] [pt] & pc = tgt if any flag 0\\
		\midrule
		rdp & rdp addr=<a> & pf = pending\_bits[addr] \\
		rdw & rdw addr=<a> lce=<l> lru\_way=<w> [src=<ra>] & produce sharers, lru info, etc.\\
		rde & rde addr=<a> lce=<l> way=<w> [src=<ra>] dst=<rd> & rd = addr, sh\_states[lce] = state\\
		wdp & wdp addr=<a> p=<0,1> & pending\_bits[addr] +/- 1\\
		clp & clp addr=<a> & pending\_bits[addr] = 0\\
		clr & clr addr=<a> lce=<l> & clear directory row\\
		wde & wde addr=<a> lce=<l> way=<w> [src=<ra>] state=<s> [state\_imm] & dir[addr, lce] = [tag, state]\\
		wds & wds addr=<a> lce=<l> way=<w> [src=<ra>] state=<s> [state\_imm] & dir[addr, lce] = […, state] \\
		gad & gad & execute GAD unit\\
		\midrule
		wfq & wfq queue [queue...] & Wait for message on queue\\
		pushq & pushq queue cmd addr=<a> lce=<l> way=<w> [src=<ra>] wp=<0,1> spec=<0,1> & push message to queue queue\\
		popq & popq queue [wp] & dequeue message, write pending bit\\
		poph & poph queue rd & capture message header\\
		specq & specq spec\_cmd addr\_sel [state] & speculation bits operation\\
		inv & inv & send invalidations\\
		\bottomrule
	\end{tabular}
	\caption{\bpbedrock{} ucode CCE Coherence ISA}
	\label{table:bedrock-isa-coherence}
\end{table*}

The ucode CCE executes a custom RISC instruction set architecture (ISA) specialized for coherence protocol processing. The ISA is divided into a \emph{Base ISA} and a \emph{Coherence ISA}. The Base ISA contains standard RISC instructions for arithmetic, branch, and data movement operations. The Coherence ISA contains instructions specialized for coherence processing, such as manipulating control flags, flag-based control flow, reading and writing the coherence directory, and sending or receiving messages. All branch instructions are tagged with a static taken/not-taken prediction bit and the branch mispredict penalty is one cycle. Table \ref{table:bedrock-isa-coherence} lists the \bpbedrock{} Coherence ISA instructions, which are divided into Flag, Directory, and Queue operations. Directory read and queue operations may take more than one cycle to execute depending on functional unit conflicts and latencies, while all other instructions execute in a single cycle.

A \emph{key contribution} of \bpbedrock{} is its specialized Coherence ISA that accelerates protocol processing by invoking coherence-specialized functional units. These instructions are critical to closing the performance gap between the ucode and FSM-based CCEs.

\subsubsection{Flag Instructions}

Flag instructions can set or clear flags, perform logic operations on pairs of flags, and make control flow decisions based on the state of a programmer-selected set of flags. The most important of these are the flag-based branch instructions (bf, bfz, bfnz, bfnot). Each flag-based branch examines a set of programmer-selected MSHR flags, encoded in a bitmask within the instruction, and branches the microcode PC to the supplied target PC if the branch condition is met. A single flag-based branch instruction is able to replace a sequence of regular branch instructions, thereby accelerating common protocol processing control flow decisions.

\subsubsection{Directory Instructions}

Directory instructions accelerate directory read, write, and processing operations by invoking the coherence directory and GAD modules. Directory way group reads require only $1 + (C/2)$ cycles to execute, compared to tens or hundreds of cycles that would be required by a general-purpose implementation of the same routine using \lstinline{for} loops. Pending bit and directory entry reads require one and two cycles, respectively. Directory writes execute in a single cycle. The GAD module executes in a single cycle, compared to a cost of tens of instructions to implement equivalent logic in general-purpose RISC code. Additionally, the flag outputs of the GAD module never need to be recomputed by the microcode program, saving many additional cycles for every flag-based branch instruction.

\subsubsection{Queue Instructions}

Queue instructions enable efficient sending and receiving of coherence protocol and memory messages. The ucode CCE is able to send and receive messages with a cost of one cycle per message header or data beat. The invalidate (inv) instruction further accelerates coherence protocol processing by invoking a small hardware-implemented state machine within the ucode CCE's message unit to efficiently send invalidation commands to all caches with a Shared (S) copy of the specified cache block at a rate of one message per cycle. A general-purpose RISC routine for invalidations would require at least a few instructions per invalidation sent if executed in a tight \lstinline{for} loop.

\subsubsection{Programming the CCE}

The CCE is programmed at the microcode level. A custom assembler applies a limited set of instruction transformations to map available software pseudo-ops into hardware-implemented microcode instructions. \bpbedrock{}'s MOESIF protocol microcode is only 125 instructions, which includes support for uncacheable access to both cacheable and uncacheable memory and system initialization.

\subsection{Instruction Fetch and Predecode}

The Instruction RAM and Predecode modules comprise the Fetch Stage of the ucode CCE. The instruction RAM unit contains the microcode instruction memory and logic to determine the next microcode program counter (PC). The predecode module examines the just fetched instruction to detect branches and provides a predicted PC to the fetch logic. A stall in the Execute stage halts instruction fetch, and a branch mispredict squashes the just fetched instruction while redirecting fetch to the correct PC.

\subsection{Instruction Decode and Stall}

The Instruction Decode unit expands the current narrow microcode instruction into a wider decoded instruction containing control signals for the Execute stage's functional units.

The Instruction Stall unit detects functional unit hazards and message stalls due to busy or empty networks. It outputs a stall signal that is routed to the Fetch stage and the instruction decoder, causing the current instruction to replay in the following cycle. Functional unit hazards occur when the ucode engine and message unit access the same resource.

\subsection{Source Select and Arbitration}

The source select module routes operands to the ucode CCE's functional units as specified by he current instruction. Source operands may come from the GPRs, MSHR, inbound messages, or directory outputs.

The arbitration unit controls access to the coherence directory, pending bits write port, and speculative bits read port. In a given cycle, each of these three resources may be used by either the microcode instruction or the message unit. The message unit has priority over the ucode engine for each resource, causing the ucode engine to stall when it loses arbitration, which helps guarantee deadlock freedom in the coherence protocol.

\subsection{Register File}

The Register File stores the CCE's Miss Status Handling Register (MSHR), eight 64-bit general purpose registers (GPRs), a coherence state register, and an auto-forward control register. The coherence state register holds a default coherence state that can be used as a source operand for coherence and memory commands. The auto-forward control register is a single bit register that controls whether the ucode CCE's message unit will automatically process memory response messages. It is set by default, but can be disabled via the microcode.

\subsection{Functional Units}

The ucode CCE includes functional units tailored for both general purpose and coherence specialized execution. The Coherence Directory, Pending Bits, Speculative Bits, and GAD modules are all re-used without modification from the FSM CCE design and are described in Section \ref{sec:fsm_cce}.

\subsubsection{ALU and Branch}

The Arithmetic Logic Unit (ALU) has a 64-bit datapath and supports add, subtract, shift, and bitwise operations. The hardware ALU is very simple, and many operations available to the programmer are supported at the software level as pseudo-ops using assembler transformations.

The Branch unit resolves branch operations and validates the Fetch stage's speculative fetch predictions. The result of the branch is compared to the branch prediction made in the Fetch stage to determine if a mispredict occurred. Mispredicts redirect the fetch stage to the resolved PC and result in a single cycle bubble in the execute stage.

\subsubsection{Message}

The Message unit is responsible for sending and receiving all coherence and memory messages. It can write the pending bits, read the speculative bits, and write the coherence directory. It also contains a memory credit flow counter that limits the number of outstanding memory commands. The message unit has two state machines that process memory response messages and send or receive messages as directed by the microcode program. The memory response state machine is identical to the one found in the FSM CCE. However, it can be disabled via the microcode program by clearing the auto-forward register in the register file. The other state machine sends and receives messages based on the currently executing instruction. This FSM also implements the specialized invalidation routine logic that can issue one invalidation per cycle.

\subsection{Request Processing} \label{sec:ucode_cce_req}

\begin{figure}[t]
	\centering
	\includegraphics[width=\linewidth]{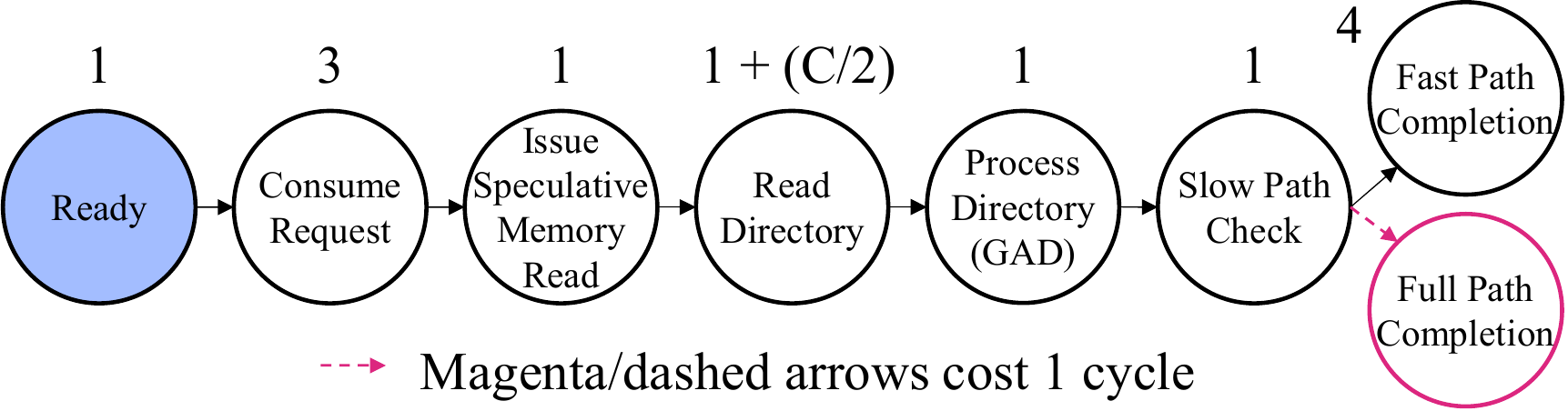}
	\caption{MOESIF Microcode Processing Flow - Initial and Fast Path}
	\label{fig:ucode-cce-moesif-front}
\end{figure}

\begin{figure}[t]
	\centering
	\includegraphics[width=\linewidth]{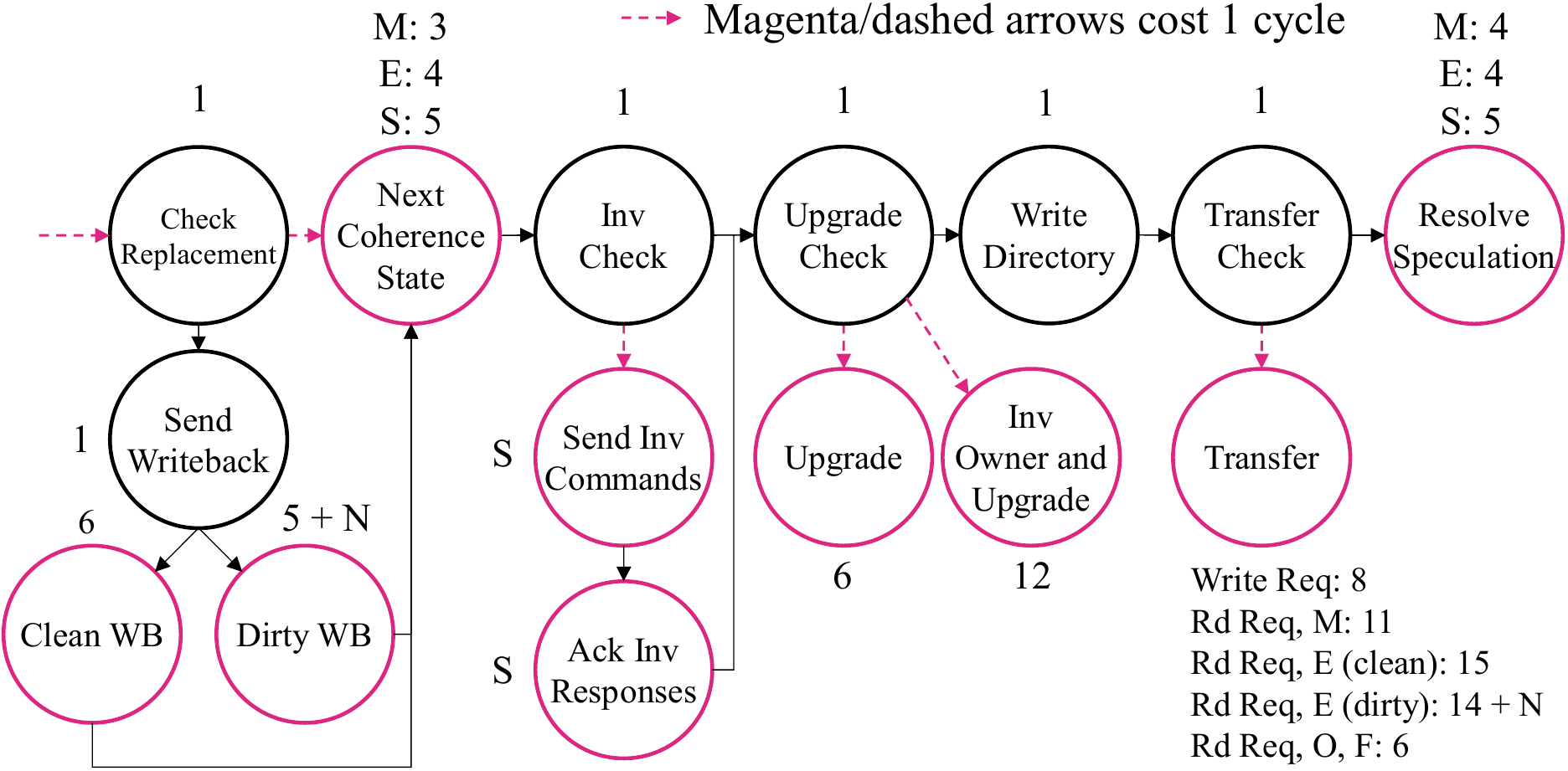}
	\caption{MOESIF Microcode Processing Flow - Full Path}
	\label{fig:ucode-cce-moesif-full}
\end{figure}

The ucode CCE's microcode programs implement an execution flow similar to the FSM CCE. Figures \ref{fig:ucode-cce-moesif-front} and \ref{fig:ucode-cce-moesif-full} show the MOESIF protocol microcode processing flow for cacheable requests. Each circle represents one or more microcode instructions and is called a subroutine, and each subroutine is labelled with its execution latency in cycles. Transitions between states that are black/solid cost zero cycles and those that are magenta/dashed cost one cycle, which indicate a branch mispredict. States outlined in magenta require more than one cycle to execute.

Figure \ref{fig:ucode-cce-moesif-front} shows the initial subroutines executed by the MOESIF protocol for all coherence requests to cacheable memory. The microcode is optimized to handle load requests to blocks in the Invalid state, which is called the \emph{Fast Path}. All other requests are handled by the \emph{Full Path}. A fast path request is fulfilled by a memory access, which is issued speculatively before the directory read occurs. The directory read latency is $1 + (Cores/2)$ cycles. Following the directory read, a single cycle is required to confirm the request does not require the full path for processing before finalizing the past path processing.

The ucode CCE MOESIF protocol continues processing non-fast path requests using the flow shown in Figure \ref{fig:ucode-cce-moesif-full}. These requests either target a block that is already cached somewhere in the system, require a cache block replacement at the requesting LCE, or are write requests. The total execution latency of all requests can be computed directly from the processing flow diagrams by adding the cost per state visited plus one cycle per branch mispredict transition (magenta/dashed arrow). The \emph{Compute Next Coherence State} and \emph{Resolve Speculation} subroutines require a variable number of cycles to execute, depending on the coherence state that will be supplied to the requesting cache. Invalidations are offloaded to the message unit and require a single cycle per invalidation sent or received, for a total of $2*S$ cycles to complete all invalidations. Writebacks during cache block replacement or from a transfer require N cycles to forward the N cache block data beats from the LCE response to the memory network. The \emph{Transfer} subroutine occupancy depends both on the type of request (read or write) and the coherence state of the block at the directory. Request processing completes after performing an Upgrade, Transfer, or Resolving Speculation (indicated by states with no out transitions).

\section{Evaluation} \label{sec:evaluation}

\subsection{Protocol Verification}

The \bedrock{} cache coherence protocol is similar to, yet subtly different from, commonly understood directory-based protocols \cite{mcm-primer-v2, sweazey1986, papamarcos1984, tang1976, censier1978}. Therefore, it is important that the protocol itself is shown to be correct. \bedrock{}'s MESI protocol has been verified correct using CMurphi~\cite{cmurphi}, an improved version of the Murphi~\cite{murphi} model checking framework, for a system with eight coherent caches and a single coherence directory. \bedrock{}'s CMurphi description models a single cache block, a single coherence directory, and unordered networks. These assumptions are valid because, by definition, cache coherence is constrained to a single memory location, \bedrock{}'s coherence directories operate independently from one another in a multi-directory system, and every cache block is managed by a single directory.

\subsection{\bedrock{} Directory Overhead}

\begin{table}[t]\centering
\begin{tabular}{@{}L{0.12\linewidth}C{0.2\linewidth}C{0.2\linewidth}C{0.3\linewidth}@{}}\toprule
Caches & \bpbedrock{} & Complete & Coarse (8-bits) \\
\midrule
2 & 6.25\% & 7.81\% & 7.81\% \\
4 & 6.25\% & 7.81\% & 9.38\% \\
8 & 6.25\% & 9.38\% & 9.38\%\\
16 & 6.25\% & 10.94\% & 9.38\%\\
32 & 6.25\% & 14.06\% & 9.38\%\\
64 & 6.25\% & 20.31\% & 9.38\%\\
\bottomrule
\end{tabular}
\caption{Coherence Directory Overhead Comparison}
\label{table:directory-overhead}
\end{table}

\bpbedrock{} utilizes a standalone, duplicate tag coherence directory. However, this is not the only choice of directory organization that could have been made. Therefore, we investigate the coherence directory storage overhead of \bpbedrock{}'s directory and compare it to the overhead of standalone complete and coarse directories \cite{mcm-primer-v2, weber1992}. The analysis assumes 8-way associative L1 caches with 64 sets and 64-byte cache blocks, which are the default for \bpbedrock{}, with private instruction and data caches per core. The system uses 28-bit physical address tags and 3-bit coherence states. Table \ref{table:directory-overhead} shows the results of the overhead analysis. \bpbedrock{}'s duplicate tag directories have a constant storage overhead of 6.25\%, which is less than both the complete and coarse directories. This property is greatly beneficial for the physical design of \bpbedrock{}, where a slice of the coherence directory is instantiated on each multicore tile and the multicore is constructed by instantiating tiles in a 2D mesh. The constant overhead results in a fixed directory size per tile, regardless of core count, which enables the use of a hierarchical, tile-based backend design flow for ASIC implementations.

\subsection{CCE Request Processing Occupancy}

\begin{table*}[t]\centering
	\begin{tabular}{@{}L{0.1\linewidth}C{0.1\linewidth}C{0.12\linewidth}C{0.29\linewidth}C{0.29\linewidth}@{}}\toprule
		Request & LCE State & Directory State & FSM CCE Occupancy (cycles) & ucode CCE Occupancy (cycles) \\
		\midrule
		Read Excl & \multirow{2}{*}{I} & \multirow{2}{*}{I} & \multirow{2}{*}{$8 + (C/2)$} & $12 + (C/2)$ \\ 
		Read NE & & & & $26 + (C/2)$ \\ 
		\midrule
		\multirow{5}{*}{Read} & \multirow{5}{*}{I} & S & $8 + (C/2)$ & $26 + (C/2)$ \\ 
		& & E (clean) & $9 + (C/2)$ & $36 + (C/2)$ \\ 
		& & E (dirty) & $9 + (C/2) + N$ & $35 + (C/2) + N$ \\ 
		& & M & \multirow{2}{*}{$9 + (C/2)$} & $32 + (C/2)$\\ 
		& & O, F && $27 + (C/2)$\\ 
		\midrule
		\multirow{4}{*}{Write} & \multirow{4}{*}{I} & I & $8 + (C/2)$ & $23 + (C/2)$ \\ 
		& & S & $8 + (C/2) + (2*S)$ & $24 + (C/2) + (2*S)$ \\ 
		& & E, M & $9 + (C/2)$ & $27 + (C/2)$ \\ 
		& & O, F & $9 + (C/2) + (2*S)$ & $28 + (C/2) + (2*S)$ \\ 
		\midrule
		\multirow{2}{*}{Write} & \multirow{2}{*}{S} & S & $9 + (C/2) + (2*(S-1))$ & $24 + (C/2) + (2*(S-1))$ \\ 
		& & O, F & $9 + (C/2) + (2*S)$ & $30 + (C/2) + (2*S)$ \\ 
		\midrule
		Write & O, F & O, F & $9 + (C/2) + (2*S)$ & $24 + (C/2) + (2*S)$ \\
		\bottomrule
	\end{tabular}
	\caption{\bpbedrock{} CCE Request Occupancy (MOESIF)}
	\label{table:cce-occupancy-moesif}
\end{table*}

The \bpbedrock{} coherence engines both implement the \bedrock{} MOESIF cache coherence protocol. The FSM CCE is a direct hardware implementation of the protocol's request processing flow, while the ucode CCE executes a microcode program implementing the protocol and provides flexibility to add additional processing and functionality to protocol processing. The throughput and latency of cache coherence engines are important to overall system performance, therefore we first analyze each coherence engine's request processing occupancy.

Table \ref{table:cce-occupancy-moesif} presents the request processing occupancy for both coherence engines. Processing occupancy, given in cycles, is the number of cycles required in a best-case, no-contention execution to process a coherence request. Three constants are used in the processing occupancy computations: C is the number of cores in the multicore processor, N is the number of data beats required to send a full cache block across the coherence network data channels, and S is the number of caches holding a block in the Shared (S) coherence state, called the sharers. The data presented are the number of cycles that the coherence engine is busy processing a single request, and processing occupancy are derived directly from Figures \ref{fig:fsm-cce-req-fsm}, \ref{fig:ucode-cce-moesif-front}, and \ref{fig:ucode-cce-moesif-full}. The numbers presented assume that a cache block eviction (replacement) is not required. Occupancy provides insight into the maximum achievable throughput of the coherence engine designs. The request occupancy does not include the time required to process memory responses, which are handled by a separate state machine in both designs that operates concurrent to request processing. Network time is also excluded as the time for messages to transit networks is the same for both designs.

\subsubsection{FSM CCE Occupancy}

The FSM CCE has a base request processing occupancy of $7 + (C/2)$ cycles, incurred by all requests, as it moves from Ready through Write Next State in Figure \ref{fig:fsm-cce-req-fsm}. During this initial processing, the request is consumed, the directory is read and processed, and the directory entry for the requesting cache is updated with the final next state for the block. Then, depending on the specific request and state of the target block in the system, the FSM executes only those steps required to complete the transaction. The diamonds in Figure \ref{fig:fsm-cce-req-fsm} indicate control flow decisions, but have a cost of zero cycles. The key performance advantage of the FSM-based design is that control flow decisions are effectively free; in any given state, the next state is computed concurrently with the protocol processing occurring in the state. Thus, after executing the initial processing, the added cost to complete a request is simply the cost of the remaining states visited. The worst-case request, in terms of occupancy, is a write request to a block in the O or F state, which is owned by a single cache but shared by many caches and may be present in every single cache in the system. A cache block replacement adds either two or $1+N$ cycles of processing time for clean and dirty blocks, respectively.

\subsubsection{ucode CCE Occupancy}

The ucode CCE incurs execution overheads relative to the FSM-based CCE primarily due to its inability to execute protocol processing and control flow in the same instruction and the fact that each control flow decision requires a separate instruction. As described in Section \ref{sec:ucode_cce_req}, the MOESIF microcode program includes a fast path (Figure \ref{fig:ucode-cce-moesif-front} to process regular reads for blocks in the Invalid state. This path has an execution overhead of only four cycles compared to the FSM-based coherence engine. The fast path is effectively a single basic-block of microcode, and therefore can be executed at a rate matching that of the FSM-based engine. However, all other requests must branch to the full path, shown in Figure \ref{fig:ucode-cce-moesif-full}, which is capable of performing replacements, invalidations, and cache to cache transfers. The base occupancy for both paths is only one cycle greater than the FSM-based engine at $8 + (C/2)$ cycles. Requests processed by the full path have occupancy overheads between 15 to 25 cycles. Significant overheads are incurred for subroutines that require multiple control flow decisions. In particular, determining the proper next coherence state for the block, resolving the outcome of the speculative memory access, and initiating cache to cache transfers all add significant latency to request processing. A cache block replacement adds either seven or $6+N$ cycles of processing time for clean and dirty blocks, respectively.

\subsubsection{Discussion}

Designing a microcode programmable coherence engine with processing occupancy equivalent to an FSM-based design remains an open challenge. The ucode CCE's contribution to this problem is its use of coherence-specific ISA extensions to efficiently offload common coherence protocol operations, such as reading and processing the directory and performing invalidations, to specialized functional units. Despite the control flow overheads experienced by the ucode CCE, processing occupancy overheads are limited to tens of cycles through the use of coherence-specific instructions. The coherence directory hardware accelerates directory operations, replacing a potentially expensive loop-based microcode routine requiring tens to hundreds of cycles with a comparatively inexpensive fixed-latency execution. Similarly, constructing and sending coherence messages are executed by the specialized message unit at a rate of one cycle per header or data beat. A reasonable software implementation might require tens of cycles per message send or receive using memory-mapped messaging queues. The ucode CCE's inclusion of coherence-specialized functional units points to a promising path forward for programmable coherence engines that are specialized for protocol processing yet flexible enough to enable unique system features.

\subsection{Area, Utilization, and Timing} \label{sec:evaluation_area}

\begin{table}[t]\centering
\begin{tabular}{@{}L{0.12\linewidth}C{0.25\linewidth}C{0.25\linewidth}C{0.2\linewidth}@{}}\toprule
Design & Component & Resource & Overhead \\
\midrule
\multirow{3}{*}{ASIC} & Multicore & \multirow{3}{*}{Die Area} & $4.08\%$ \\
& Tile & & 4.28\% \\
& CCE & & 31.08\%\\
\midrule
\multirow{5}{*}{FPGA} & \multirow{2}{*}{Multicore} & Logic LUTs & 6.32\% \\
& & BRAM & 1.54\%\\
\cmidrule{2-4}
& \multirow{2}{*}{Tile} &  Logic LUTs & 7.08\% \\
& & BRAM & 1.54\%\\
\cmidrule{2-4}
& \multirow{2}{*}{CCE} &  Logic LUTs & 66.19\%\\
& & BRAM & 1 per CCE\\
\bottomrule
\end{tabular}
\caption{\bpbedrock{} ucode CCE Resource Overheads}
\label{table:fpga}
\end{table}

\blackparrot{}, including \bpbedrock{}, has been silicon validated using GlobalFoundries 12nm FinFET process and FPGA validated in an 8-core configuration for each coherence engine using a Xilinx Ultrascale+ VCU128 development platform\cite{xilinx-ultrascale}. Table \ref{table:fpga} provides area and resource utilization overheads for ASIC and FPGA-based designs. The overheads listed are normalized to designs using the FSM-based coherence engine. The more efficient ASIC implementations show the introduction of programmability into the coherence system comes at a small area cost of only 4.08\% extra die area for the entire multicore and a 4.28\% increase per \blackparrot{} Tile. Each \blackparrot{} Tile comprises a \blackparrot{} core, its 32 KiB L1 D\$ and I\$, a 64 KiB slice of the distributed L2 cache, the on-chip networks and routers to connect tiles, and an instance of the \bpbedrock{} coherence engine and directory. The L2 cache acts as a memory-side buffer and does not participate in the coherence protocol. All SRAM macros are hardened in the ASIC flow, and the multicore is a 2-D mesh of Tiles with minimal additional surrounding logic. These area overheads are largely due to the addition of a microcode instruction SRAM. In the FPGA implementations, the logic utilization increases by only 6.32\% and 7.08\% for the entire multicore and per tile, respectively, when using the ucode CCE. Each programmable CCE additionally requires a single 18 Kib block RAM resource, which amounts to a 1.54\% increase in 18 Kib block RAM resources\footnote{36 Kib block RAMs are counted as two 18 Kib block RAMs for this analysis.}. Additionally, both coherence engine implementations meet the same design target frequency in both the ASIC and FPGA implementations.


\subsection{Splash-3 Performance} \label{sec:evaluation_splash3}

\begin{figure}[t]
	\centering
	\includegraphics[width=\linewidth]{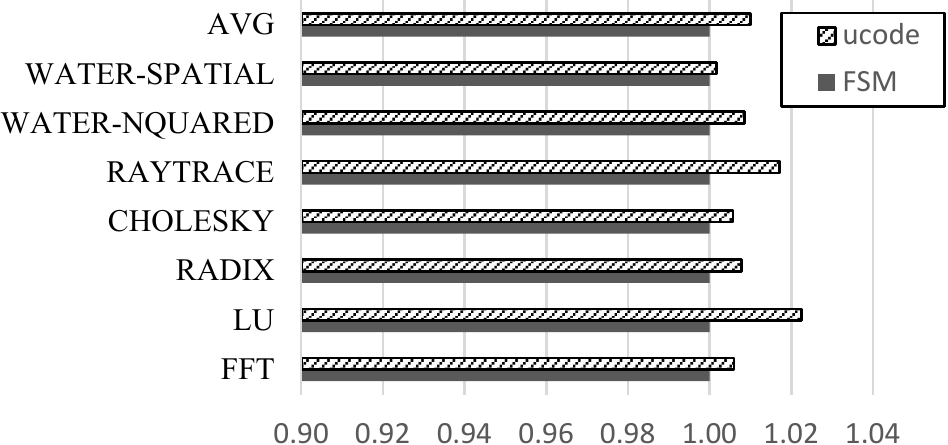}
	\caption{8-core \bpbedrock{} Splash-3 Normalized Execution Time}
	\label{fig:splash3}
\end{figure}

To compare the impact of coherence engine design on system performance we ran a collection of benchmarks from the Splash-3 \cite{sakalis2016} suite on FPGA-based 8-core \bpbedrock{} systems with the FSM-based and microcode programmable coherence engines. The benchmarks were compiled for the RISC-V ISA using \lstinline{gcc} targeting a Linux environment, and invoked to execute on all 8 available processor cores. The FFT, LU, RADIX, and CHOLESKY programs are smaller kernel programs, while the remaining three programs are larger application programs. We refer readers to \cite{sakalis2016} and \cite{woo1995} for more details on the synchronization and memory characteristics of these programs. Wall-clock execution time for all benchmarks ranged between tens of seconds and tens of minutes.

The \bpbedrock{} FPGA designs instantiate an 8-core \blackparrot{} multicore with 32 KiB L1 instruction and data caches, a 512 MiB shared L2 cache, and 2 GiB of HBM-based main memory. The multicore instance was clocked at 50 MHz. The programs were run in a Linux-based OS environment constructed using BuildRoot \cite{buildroot} and Busybox \cite{busybox} with Linux kernel v5.15 \cite{linux} and OpenSBI v1.0 \cite{opensbi}.

Figure \ref{fig:splash3} shows total execution time measured using the \lstinline{time} utility for each benchmark, averaged over three runs, and normalized to the FSM-based multicore design. Our experiments show that the microcode programmable coherence engine is within 1\% of the hardware-based FSM's performance on average, and only 2.3\% slower at worst. Despite having a 15 to 25 cycle best-case processing occupancy overhead, the ucode CCE multicore experiences a very small performance slowdown. Intuitively, this result makes sense as any program with reasonably good cache utilization and low miss rates will only invoke the coherence system on a cache miss. If misses are infrequent, the overall impact on system performance will be small, which follows directly from the standard average memory access time computation.

This result indicates a promising path forward for further exploration of programmable coherence engines. Careful design of the protocol processing paths can keep a programmable coherence engine competitive with a fixed-function engine, while the flexibility of a programmable system can unlock exciting new system features.

\section{Related Work} \label{sec:related}

\bpbedrock{} is most closely related to prior research studying programmable cache coherence systems. Programmable coherence systems have been explored at many levels. At one extreme, software-managed cache coherence~\cite{schoinas1994, lee2008, lee2011, pai2012} and systems without any hardware coherence support~\cite{IntelSCC,gries2011} rely on the system or application programmer to implement correct coherence mechanisms -- no easy feat! \bpbedrock{} provides flexibility to the system designer through its programmable coherence engine, but does not require the application developer to manage coherence.

The benefits of programmable controllers are evaluated in~\cite{michael1997, michael1999}, and controller designs include the MAGIC node controller~\cite{kuskin1994,heinrich1994,kuskin1997-phd,heinrich1998-phd}, Wisconsin Typhoon~\cite{reinhardt1994}, Sun Microsystems S3.mp~\cite{nowatzyk1994}, and the Piranha chip multiprocessor~\cite{barroso2000}. Piranha and S3.mp use microprogrammed coherence engines that are more specialized than \bpbedrock{}, Typhoon uses an off the shelf commodity processor as the protocol engine, and MAGIC uses a customized MIPS-based RISC processor as the node controller.

\begin{table}[t]\centering
	\begin{tabular}{@{}L{0.32\linewidth}C{0.29\linewidth}C{0.29\linewidth}@{}} 
        \toprule
		Property & MAGIC & \bpbedrock{} \\
        \midrule
		Base ISA & MIPS & Custom RISC \\
		\multirow{3}{*}{ISA Extensions}  & Bitfield Op & Directory Rd/Wr \\
		& Set/Test Bit  & Flag Op \\
		& Tx/Rx Message & Tx/Rx Message \\
		GPRs & 32 x 64-bit & 8 x 64-bit \\
		Data \$ & 32 KiB off-chip & none \\
		Instruction \$ & 16 KiB on-chip & 1.5 KiB \\
		Data Buffers & 2 KiB 6-port SRAM & none \\
		Directory Memory & none & 3.625 KiB \\
		Protocol Agnostic & yes & no \\
		Message Passing & yes & no \\
		Coherence Type & Distributed Directory & Distributed Directory \\
		Coherence Domain & Inter-node & Multicore \\
		Coherence Model & All memory blocks & Only cached blocks \\
		HW Address Translation & no & no \\
		Interrupts & no & no \\
		Open Source & no & yes \\
        \bottomrule
	\end{tabular}
	\caption{Architectural Properties of \bpbedrock{} and MAGIC}
	\label{table:magic-properties}
\end{table}

\begin{table}[t]\centering
    \begin{tabular}{@{}L{0.32\linewidth}C{0.29\linewidth}C{0.29\linewidth}@{}} 
    	\toprule
    	Operation & \bpbedrock{} & MIPS (MAGIC) \\
    	\midrule
    	Directory Read & $2 + C/2$ & $20*C$ \\
    	Invalidation & $2+(2*S)$ & $15*S$ \\
    	Branch & 1 & 1 \\
    	Flag Branch & 1 & -- \\
    	\bottomrule
    \end{tabular}
    \caption{Selected operation latency in cycles. C is the number of cores and S is the number of sharer caches.}
    \label{table:magic-dir-latency}
\end{table}

\begin{table}[t]\centering
	\begin{tabular}{@{}L{0.28\linewidth}C{0.30\linewidth}C{0.30\linewidth}@{}} 
		\toprule
		Request (Directory State) & \bpbedrock{} & MIPS (MAGIC) \\
		\midrule
		Read (I) & 16 & 184 \\
		Read (S) & 30 & 184 \\
		Read (M) & 36 & 219 \\
		Write (I) & 27 & 184 \\
		Write (S) & $28 + (2*S)$ & $184 + (15*S)$ \\
		Write (M) & 31 & 198 \\
		\bottomrule
	\end{tabular}
	\caption{Request Occupancy in cycles, assuming 8-cores and an invalid block at the requester. The coherence state in parentheses indicates the state of the block at the directory. S is the number of sharer caches.}
	\label{table:magic-dir-occupancy}
\end{table}

\bpbedrock{} is most similar to MAGIC in that both designs are effectively small, specialized integer-only RISC ISA engines. Unlike MAGIC, which is designed as a generic protocol processor, \bpbedrock{}'s programmable engine is designed to efficiently implement the \bedrock{} coherence protocol while enabling unique system- and application-specific functionality via programmable routines executing alongside protocol processing. \bpbedrock{} is not designed to support arbitrary coherence protocols or shared memory solutions. Table \ref{table:magic-properties} compares the architectural properties of MAGIC and \bpbedrock{}.
\bpbedrock{} includes dedicated directory storage and a microcode instruction memory instead of general purpose instruction and data caches. Both designs use specialized RISC instruction sets with similar extensions for bit manipulations and message send and receive operations, however, \bpbedrock{} also includes specialized instructions for reading and processing the coherence directory and to perform efficient flag based control flow. Neither \bpbedrock{} or MAGIC supports virtual memory or interrupts. Tables \ref{table:magic-dir-latency} and \ref{table:magic-dir-occupancy} provide quantitative comparisons between \bpbedrock{} and a MIPS-based protocol processor like MAGIC. Table \ref{table:magic-dir-latency} compares the latency of selected directory operations such as reading and processing a duplicate tag directory, issuing invalidations, and control flow operations. \bpbedrock{}'s specialized functional blocks enable highly efficient coherence directory reads, while a MIPS-based protocol processor such as MAGIC requires a significant number of instructions to execute the same operation. Likewise, \bpbedrock{}'s specialization allows it to issue one invalidation command per cycle and consume one response per cycle, whereas a MIPS-based processor would require executing these routines as tight loops with approximately 10 instructions per send or receive operation. Table \ref{table:magic-dir-occupancy} shows the processing occupancy in cycles. at the coherence directory for common requests. The table assumes the requesting cache does not have a valid copy of the block, which is currently in the coherence state listed in parentheses at the directory. \bpbedrock{}'s specialized logic for reading and processing the directory, issuing invalidations, and executing control flow decisions based on the coherence-specific MSHR control flags give \bpbedrock{} a significant advantage over the MIPS-based execution of MAGIC.

Within the RISC-V and open-source hardware communities, \bpbedrock{} is related to OpenPiton~\cite{balkind2016}, Rocket \cite{asanovic2016rocket}, BOOM \cite{celio2015berkeley}, Chipyard \cite{chipyard}, and Ariane \cite{zaruba2019}. Each of these processors or platforms can be used to create an open-source multicore processor, however Rocket, BOOM, and Ariane are all focused more on individual core design rather than multicore system design. Chipyard is a generator framework that is capable of creating SoCs with multiple cache-coherent cores. \bpbedrock{} differs from all of them in that it focuses on the design and integration of the cache coherence system into the multicore design, and the complete implementation is in industry-standard SystemVerilog. Both OpenPiton and \bpbedrock emply directory-based cache coherence, but OpenPiton contains only a fixed-function coherence engine and embeds the directory information in the L2 cache. \bpbedrock{} includes a programmable coherence engine that is decoupled from the L2 cache, allowing for varied L2 cache implementations. Both designs are influenced by the OpenSparc T1 architecture~\cite{opensparct1}, with OpenPiton using modified OpenSparc T1 cores and \bpbedrock{} drawing inspiration from the T1's cache coherence protocol. To the best of our knowledge, \bpbedrock{} is the first programmable coherence engine for a modern multicore design.

\section{Conclusion} \label{sec:conclusion}

This paper presents \bpbedrock{}, the open-source cache coherence protocol and system implemented within the \blackparrot{} 64-bit RISC-V multicore processor. \bpbedrock{} implements the \bedrock{} directory-based cache coherence protocol and includes two different open-source coherence protocol engines, one FSM-based and the other microcode programmable. \bpbedrock{} has been validated in both ASIC- and FPGA-based implementations. Analysis shows that the programmable coherence engine increases die area by only 4\% in an ASIC process and increases logic utilization by only 6.3\% on FPGA with one additional block RAM added per ucode CCE. FPGA-based experiments with the Splash-3 benchmark suite show that the programmable controller's performance is only 1\% slower than the FSM-based CCE on average (2.3\% worst-case). These results show that it is possible to introduce programmability into the cache coherence system at a reasonable complexity and peformance cost, encouraging further exploration into the benefits of programmable coherence systems.

\bibliographystyle{IEEEtranS}
\bibliography{refs}

\end{document}